\documentstyle[prl,twocolumn,aps]{revtex}
\def\be{\begin{equation}}
\def\ee{\end{equation}}
\def\bea{\begin{eqnarray}}
\def\eea{\end{eqnarray}}
\begin{document}
\draft

\title{Iterative maps with hierarchical 
clustering for the observed \\ 
scales of astrophysical and cosmological 
structures}
\author{
Salvatore Capozziello$^{*}$,
Salvatore De Martino$^{\S }$,
Silvio De Siena$^{\S }$, \\
Francesco Guerra$^{\ddag}$ and
Fabrizio Illuminati$^{\S }$}

\address{
$^*$ Dipartimento di Scienze Fisiche 
``E. R. Caianiello", 
Universit\`a di Salerno, 
and INFN, Sez. di Napoli, \\
G. C. di Salerno, via S. Allende, 
I--84081 Baronissi (SA), Italy \\ 
$^{\S}$ Dipartimento di Fisica, 
Universit\`a di Salerno, INFN, Sez. di
Napoli, G. C. di Salerno, \\
and INFM, Unit\`a di Salerno, 
via S. Allende, I--84081 Baronissi (SA), 
Italy \\ 
$^{\ddag}$ Dipartimento di Fisica, 
Universit\`a di Roma ``La Sapienza'',
and INFN, Sez. di Roma, \\
Piazzale A. Moro, I--00185 Roma, Italy}

\date{\today}

\maketitle

\begin{abstract}
We compute the order of magnitude 
of the observed 
astrophysical and cosmological scales
in the Universe, from neutron 
stars to superclusters of galaxies, 
up to, asymptotically, the observed 
radius of the Universe. 
This result is obtained by 
introducing a recursive scheme
that yields a rapidly convergent 
geometric sequence, 
which can be described as a 
hierarchical clustering.
\end{abstract}

\vspace{0.4cm}


The theoretical understanding
of the observed scales, sizes
and dimensions of
aggregated structures in the
Universe (stars, galaxies, clusters,
etc.) is a long--standing open problem
in astrophysics and cosmology.

In this letter, we introduce a 
geometric description
which accounts with accuracy 
for the order of magnitude,
and provides a rapidly converging 
succession for the observed
length and mass scales of 
the main astrophysical and
cosmological structures.

We base our analysis on
the {\it granularity} of the Universe,
in the sense that nucleons 
are taken as
the building blocks of 
the observed stable
cosmological aggregates.
We further stipulate that
only the gravitational interaction
accounts for the gross features
(sizes, masses, number of components)
of the observed cosmological and 
astrophysical aggregates.
It is worth noting that, 
if this hypothesis can be
considered obvious when one 
looks at large scale
cosmological structures, 
it appears rather nontrivial
for astrophysical ones,
e.~g., stars, in which
other interactions of 
nongravitational nature
play a relevant role. 
However, as will be discussed
and clarified in the following,
our scheme allows to identify
the sizes of 
those astrophysical structures,
like neutron stars and planetary
systems, where gravity
is the only overall effective
interaction.

The iterative 
scheme introduced in the
present paper provides the
scales of astrophysical and 
cosmological structures 
as a hierarchical
sequence of ``close--packed''
aggregates of increasing
size in a spatially 
flat Universe, i.~e. with
a density parameter $\Omega \simeq 1$.
This finding is strongly supported
by the recent evidences 
coming from the BOOMERANG experiment 
\cite{melchiorri}, and
explains the good agreement
with the observational data of our
theoretical scheme
which assumes a space--like sheet 
embedded in a space--time cosmological
manifold. In fact, the
order of magnitude of the limiting 
size in our
iterative procedure, i.e. the
``observed radius of the
Universe'', will turn out to be 
$\cong 10^{26} cm$,
which actually coincides 
with the observed
distance that can be measured without
introducing second order 
corrections to the 
Hubble law \cite{peebles}.

The scheme basically consists of 
successive iterations of two 
alternating
physical mechanisms.
The first mechanism is suggested 
by the tendency
towards collapse for a system 
of gravitationally interacting
bodies, due to the long range nature 
of the gravitational force.
This tendency to a three--dimensional 
(3--d) close packing
is however opposed by the
relativistic constraint of 
the maximum attainable
(critical) gravitational
energy for the system (essentially, 
the rest mass).
This constraint
leads to a ``critical 3--d close 
packing'' that 
singles out a minimal scale 
of aggregation.

The critical 3--d close packing yields,
for the smallest aggregate,
a spatial gravitational
energy density exceedingly large 
with respect to a minimum mean
spatial gravitational energy density,
which will be defined below
(here and in the following,
if not otherwise specified, we take
the gravitational energies and energy
densities always in modulus).
This latter quantity 
will turn out to be the
spatial energy density associated  
to the asymptotic length
scale in our iterative scheme
(mean energy density of 
the observed Universe).

Therefore, in the second 
step of the iteration,
the mass of the smallest 
aggregate is redistributed
on a larger spatial scale 
in such a way to bring
the mean energy density of 
the new aggregate to
coincide with the mean 
energy per unit
volume of the observed Universe.
This condition implies 
a proportionality between
the total mass of the 
new aggregate and the
square of the new spatial 
radius, and is
equivalent to a 
two--dimensional (2--d) close
packing.

The mass distribution thus obtained
is confirmed, at least on large scales,
by the statistical analysis on cosmological
data catalogues performed in recent years
\cite{pietronero}, in particular on
the statistical correlation among
galaxies.
Observing that, on scales smaller
than $10^{25} \div  10^{26} cm$ the matter
distribution cannot be considered homogeneous,
the authors in Ref. \cite{pietronero}
assume a fractal behavior that yields
a statistical density--density correlation
decreasing with the inverse of the length scale.
This result implies that the aggregates'
masses must be proportional to the
square of the aggregates' radii.

The recursive scheme is implemented by
iterating the two different processes
of aggregation in alternating
order: critical 3--d close
packing of a system made of second--step aggregates,
enlargement of the new radius with fixed mass
in order to attain the 2--d close packing at
constant energy density, and so on.
In this way we shall obtain a sequence of length and
mass scales with rapid geometric convergence to an
asymptotic scale at which further 3--d close packing
becomes irrelevant and thus the iteration reaches
a fixed point.

In order to avoid possible sources of
confusion, we remark that our iterative 
process does not imply
that larger structures are 
generated in time from smaller ones
(rather than viceversa)
as will be clear from the following
and further discussed in the conclusions. 

We now proceed to develop explicitely the scheme.
Denote by $R_{n}$ and $M_{n}$,
respectively, the length extension and the
total mass of the
$n$--th aggregate labelled in the sense of
increasing size.
Let us define next $N_{n}$, the number of
aggregates living on the ($n - 1$)--th scale which,
in turn, form the $n$--th aggregate.
Obviously, $M_{n}$ and $N_{n}$ are not
independent objects.

Let us then introduce the quantities
$m$ and $\lambda$ associated to
the elementary constituents
(nucleons).
Here $m$ is the mass of the nucleon
(proton or neutron), in order of
magnitude: $m \cong 10^{-24} g$, while
$\lambda$ is, in order of magnitude,
the {\it spatial extension} of a nucleon,
$\lambda \cong 10^{-13} cm$. We
stress that, in this context,
$\lambda$ is simply the linear dimension
of the space region forbidden to 
penetration, due to the
presence of a nucleon, as determined, e.g., 
by alpha particle
scattering experiments. In this 
framework, then,
$\lambda$ has a 
purely {\it classical} meaning,
and plays the 
role of a {\it minimum scale of length}.

Let us consider the initial,
minimal scale: $R_{0} = 
\lambda, \;  M_{0} = m,
\; N_{0} = 1$ (step zero, 
the nucleon).
In step one, the 
smallest aggregate $R_{1}, M_{1}, N_{1}$
is obtained by a critical 3--d close packing
of nucleons. This amounts, 
first, to equating
the mass density per unit volume of 
the aggregate with
that of a nucleon (3--d close packing), 
and then to
imposing the condition (criticality)
that the total gravitational
energy $G {M_{1}}^{2}/{R_{1}}$
attains the maximum value compatible with the
relativistic constraint, that is the rest energy
$M_{1}c^{2}$.
Imposing the two conditions
\be
\frac{M_{1}}{{R_{1}}^{3}} = 
\frac{m}{{\lambda}^{3}} \, ;
\; \; \; \; \; \; \;
\frac{G {M_{1}}^{2}}{R_{1}} = M_{1} \, c^2 \, ,
\label{eq:3dccp}
\ee

\noindent and solving for the radius 
$R_{1}$, and for the mass
$M_{1}$ (and for the number $N_{1}$),
we have:
\be
R_{1}^2 =\lambda \frac{(\lambda c)^2}{G m} \, ;
\; \; \; \; \;
M_{1} = N_{1} m \, ;
\; \; \; \; \; 
N_{1} =
\left( \frac{\lambda c^2}{G m} \right)^{3/2} \, .
\label{eq:data1agg}
\ee

\noindent 
Inserting numbers:
$R_{1} \cong 10^{6} cm$, 
$M_{1} \cong 10^{34} g$, and
$N_{1} \cong 10^{58}$. These data
coincide with the well known typical dimensions
of a neutron star \cite{oppenheimer}.
We note that for a neutron
star the first condition in Eq. (\ref{eq:3dccp}),
the equality of the
mass density of the star with the mass
density of the nucleon, is a well established
fact \cite{oppenheimer}.

\noindent Let us now define
\be
R \equiv \frac{(\lambda c)^2}{G m} 
= \lambda \gamma^{-1} \, ;
\; \; \; \; \; \;
\gamma \equiv \frac{\lambda}{R} 
= \frac{G m}{\lambda c^2} \, .
\label{eq:basicq}
\ee

\noindent
In terms of these two universal quantities, whose
numerical values are $R \cong 10^{26}cm$ and
$\gamma \cong 10^{-39}$, the expressions in 
Eq.~(\ref{eq:data1agg}) are recast in 
the simpler form
\be
R_{1}^2 = \lambda R \, ;
\; \; \; \; \; \;
M_{1} = \gamma^{-3/2} m \, ;
\; \; \; \; \; \;
N_{1} = \gamma^{-3/2} \, .
\label{eq:ndata1agg}
\ee

\noindent 
The length $R$ and the pure number $\gamma$
are the basic quantities in terms of which
the dimensions on all scales will be expressed.

In the second step of the iteration,
we impose that
the mass $M_{1}$ redistributes itself
on a larger radius $R_{2} >> R_{1}$,
determined by the condition that
its spatial gravitational
energy density $\rho_{2}$
takes a universal constant value
$\rho_{0}$:
\be
\rho_{2} \cong
\frac{G M_{2}^2}{R_{2}^4}
= \frac{G M_{1}^2}{R_{2}^4} =
\rho_{0} \, .
\label{eq:cenden}
\ee

\noindent Eq.~(\ref{eq:cenden})
immediately yields,
for some constant $a$,
\be
M_{2} = M_{1} = a R_{2}^2 \, .
\label{eq:2dcp}
\ee

The choice of the constant $a$ is suggested
by the fact that the nucleons are still the
fundamental objects, and therefore we postulate
the surface mass density of the aggregate
$M_{2}/{R_{2}}^{2}$, {\it to be
the surface mass density of a nucleon}:
\be
a = \frac{m}{\lambda^2}.
\label{eq:smdn}
\ee

\noindent 
The crucial choice in Eq.~(\ref{eq:smdn})
completely determines the 
recursive scheme and together
with Eq.~(\ref{eq:2dcp})
defines a 2--d close packing of nucleons.
Moreover, from Eq.~(\ref{eq:cenden}),
the spatial energy density
of the second--step aggregate 
is $\rho_{0} = G a^2$, and will
coincide with the mean spatial energy density
of the observed Universe.

Collecting Eqs.~(\ref{eq:2dcp}) and
(\ref{eq:smdn}), 
the second--step aggregate
is completely specified:
\be
R_{2}^2 = R_{1} R \, ;
\; \; \; \;
M_{2} = M_{1} \, ;
\; \; \; \;
N_{2} = N_{1} \, .
\label{eq:data2agg}
\ee

\noindent 
Inserting numbers:
$R_{2} \cong 10^{16} cm,\; M_{2} = M_{1} \cong
10^{34} g,\; N_{2} = N_{1} \cong 10^{58}$,
which correspond to the typical dimensions of
the solar system or, if one wishes, of the
interaction range of a typical star. It is
not surprising that we get the radius of
the solar system rather than the solar radius,
since our scheme summarizes the effective 
interaction on a test particle, due
to the presence of a star, through the overall
gravitational attraction and thus selects the
maximum external binding range of the 
interaction. In other words, we can state
that our scheme is capable of selecting
effective geometric lenghts (neutron stars)
and effective interaction lengths (planetary
systems). 

Consider now the spatial energy density $\rho_{1}$
of the first aggregate. From Eqs.~(\ref{eq:2dcp}),
(\ref{eq:smdn}), (\ref{eq:data2agg}), it follows that
$\rho_{1} = G a^2 {R^2}/{R_{1}^2} =
\rho_{0} \gamma^{-1}$.
We thus see, from the numerical value of $\gamma$,
that the spatial energy density $\rho_{1}$
is enormous with respect to the mean spatial energy
density $\rho_{0}$ of the second aggregate.

We now proceed to iterate at all orders the
two alternating mechanisms. The recursion
produces a sequence of odd--numbered
aggregates $R_{2k +1}$, $M_{2k + 1}$, $N_{2k + 1}$
determined by critical 3--d close packing of
even--numbered aggregates $R_{2k}$, $M_{2k}$,
$N_{2k}$, these latter in turn determined
by the 2--d close packing analogous to
condition (\ref{eq:2dcp}). Explicitely
($M_{0} = m$, and $k \geq 0$):
\[
\frac{M_{2k+1}}{R_{2k+1}^{3}} = 
\frac{M_{2k}}{R_{2k}^{3}} \, ;
\; \; \; \; 
\frac{GM_{2k+1}^{2}}{R_{2k+1}} =
M_{2k+1} \; c^{2} \, ,
\]

\be
M_{2k+2} = M_{2k+1} = aR_{2k+2}^{2} \, ;
\; \; \; \; 
N_{2k+2} = N_{2k+1} \, .
\label{eq:oecp}
\ee

The procedure is summarized by the
iterative map for the radii (linear
dimensions)
\be
R_{0} = \lambda \, ; \; \; \; \; \; \; \; \;
R_{n + 1}^2 =   R_{n} R \, ; \; \; \; \; \; \;
n \geq 0 \, ,
\label{eq:rim}
\ee

\noindent and by the iterative map for the numbers and, 
equivalently, for the masses ($N_{0} = 1$, $M_{0} = m$,
$N_{1}$ and $M_{1}$ as given in Eq.~(\ref{eq:ndata1agg}), 
and $k \geq 1$):
\[
N_{2k + 1} =
N_{2k + 2} = \frac{R_{2k + 1}}{R_{2k - 1}} = 
\gamma^{-\left(\frac{3}{4}\right)  2^{-(2k
- 1)}} \; ;
\]

\be
M_{2k + 1} = M_{2k + 2} =
N_{2k + 1} M_{2k} \; .
\label{eq:nim}
\ee

It is important to note that Eq.~(\ref{eq:rim})
expresses a relation of hierarchical clustering
between aggregates on different scales. 
From the same map we see that
$R \cong 10^{26} cm$ is the fixed point, and
thus $R = \lim_{n \rightarrow \infty} R_{n}$.
As a consequence, $R$ has the meaning of
the maximum observable length scale.

The map (\ref{eq:rim}) can be reexpressed in
the remarkable adimensional form:
\be
x_{n + 1} = x_{n}^{1/2} \, ; 
\; \; \; \; n \geq 0 \, ,
\label{eq:afmap}
\ee

\noindent
where $x_{n} = R_{n}/R$. The map (\ref{eq:afmap})
generates the relative scales from $x_{0} =
\lambda/R \equiv \gamma$
to $x_{\infty} = 1$, and can be completely
solved, yielding 
$x_{n} = (x_{0})^{2^{-n}} 
\equiv (\gamma)^{2^{-n}}$.
The fast geometric convergence to the
relevant cosmological scales is made evident
by rewriting the adimensional map
in the form $X_{n} = 2^{-n}$,
where $X_{n} \equiv \ln{x_{n}}/\ln{x_{0}}$.

The maps in Eq.~(\ref{eq:nim})
show that the aggregates on larger scales contain
fewer components in terms of aggregates defined
on the preceding scales, until, in the limit
$k \rightarrow \infty$
the sequence $N_{2k + 1}$ converges to
$N = 1$ (the Universe contains only itself).

The recursive relations in Eq.~(\ref{eq:nim})
allow also to compute the total number
$N_{nucl}$ of nucleons contained in the asymptotic
scale aggregate (i.~e. the total number of nucleons
in the observed Universe).
From Eqs. (\ref{eq:ndata1agg}) and (\ref{eq:nim})
it follows that
\be
N_{nucl} = N_{1} \; \lim_{k \rightarrow \infty}
\prod_{s=1}^{k} N_{2s + 1} =
\gamma^{-2} \; .
\label{eq:nnuc}
\ee

\noindent
Inserting the numerical value of $\gamma$,
$N_{nucl} \cong 10^{78}$, in perfect agreement
with the central value obtained from nucleosynthesis
calculations \cite{kolb}.

Finally, inserting the mass/radius relations
of Eq.~(\ref{eq:oecp})
and the recurrencies of
Eq.~(\ref{eq:rim}) into
the expression $E_{2k + 1} = GM_{2k +1}^{2}/R_{2k +1}$
of the total gravitational
energy of the ($2k + 1$)--th aggregate,
we observe that all the
odd scales (corresponding to critical 3--d
close packings) share the same {\it linear energy
density} $\rho_{0} R^2$. The mean spatial energy density
of the ($2k + 1$)--th aggregate is
thus $\rho_{0}(R/R_{2k + 1})^2$,
and can be computed in terms 
of powers of $\gamma$.
Consider now $\rho_{2k}$, $\rho_{2k+1}$,
respectively, the spatial energy
densities of the even and of the odd
aggregates. It is easy to verify that the
behavior of the spatial gravitational
energy densities as functions of 
the different scales ($k \geq 0$) reads:
\be
\ln{ \left[ \frac{\rho_{2k}}{\rho_{0}} \right] }
= 0 \, ; \; \; \; \; \;  
-\frac{1}{\ln{\gamma }}
\ln{ \left[ 
\frac{\rho_{2k+1}}{\rho_{0}} \right] }
= 2^{-2k} \; . 
\label{eq:energia}
\ee

These relations show that our
construction is based on a distribution
of the spatial gravitational energy density,
as a function of the scales, that grows
from the mean density $\rho_{0}$
of the even--numbered aggregates to
peaks for the odd--numbered aggregates,
in such a way that the peaks' heights decrease
with increasing scales. Asymptotically,
these peaks disappear, the density
ultimately acquires the universal value
$\rho_{0}$, and further 3--d close
packings become trivial. Reminding 
that we consider the moduli of
the energy densities, 
the peaks correspond to
relative minima, while $\rho_{0}$ 
is an absolute maximum.

In order of magnitude, the sequence
of length scales reads:
$10^{6} cm$, $10^{16} cm$,
$10^{21} cm$, $10^{23 \div 24} cm$, 
$10^{24 \div 25} cm$, $10^{25 \div 26} cm$.
After the sixth iteration $R_{6}$,
the fast geometric convergence
of the sequence does not allow
to single out further significant sizes but
that associated to the fixed point 
$R \cong 10^{26} cm$.
Thus, beyond the scales of the neutron
stars and of the solar system, the subsequent 
iterations yield the sizes of galaxy bulges,
giant galaxies or tight galaxy groups, 
up to clusters and superclusters. 

Some subtle conceptual issues related
to the above results deserve some comments, 
due to the apparent and puzzling
occurrence of the Planck action constant
in the cosmological context. 
We believe that the key to understand
this problem lies in the
intriguing numerical coincidence of the length
extension $\lambda$ of a nucleon, as obtained
in scattering experiments, and the Compton
length $\lambda_{c}$ appearing in quantum
electrodynamics:
\be
\lambda \cong \lambda_{c} \equiv 
\frac{\hbar}{m c} \, .
\label{eq:compton}
\ee

\noindent
This is a nontrivial point.
In fact, exploiting the identification (\ref{eq:compton})
in Eq. (\ref{eq:basicq}), we obtain
\be
R = \frac{\hbar^2}{G m^3} \, ;
\; \; \; \; \; \;
\gamma = \frac{\hbar c}{G m^2} \, .
\label{eq:dilemma}
\ee

\noindent
The expression of the observed radius of the
Universe, as given in Eq. (\ref{eq:dilemma})
is nothing but the Eddington--Weinberg relation
$G^{1/2} m^{3/2} R^{1/2} \cong \hbar$ \cite{weinberg},
while the form of $\gamma$ in Eq. (\ref{eq:dilemma})
is the gravitational fine structure constant.
Therefore it would seem that $R$ is directly determined
by the microscopic quantum background. This implication
can also be formally obtained by computing, in
semiclassical quantization, the Bohr radius of a system
formed by two nucleons mutually
interacting only via the gravitational force.
An easy calculation immediately yields $R_{Bohr} = R$.
We note that the identification of the observed
radius of the Universe through the Eddington--Weinberg
formula is at the heart of a recent proposal by
F. Calogero \cite{calogero}, which contains 
some hints that eventually lead us to
develop the scheme presented in this paper.

The numerical coincidence $N_{nucl} = \gamma^{-2}$
for the total number of nucleons, but with the
expression (\ref{eq:dilemma}) for $\gamma$,
was instead conjectured by Dirac
\cite{dirac}, with a consequent possible
role of $\hbar$ in determining 
global cosmological sizes.
We remark that, in our procedure,
we introduce the linear 
dimension of the nucleon as
the minimal length scale,
without exploring the subtle conceptual
implications of the identification
(\ref{eq:compton}). 
We thus believe that
it is this last coincidence which
should deserve a deeper understanding.

A somewhat different situation could 
hold in the case of neutron stars. In fact,
the expression for the number $N_{1}$ of nucleons
in a star given in Eq. (\ref{eq:ndata1agg}),
with the definition (\ref{eq:dilemma}) for $\gamma$,
was obtained in the seminal works of Chandrasekhar
\cite{chand} and Carter \cite{carter}, who
applied a Thomas--Fermi approximation
and considered the equilibrium
condition between the radiation pressure and the
gravitational force. In this case,
the identification (\ref{eq:compton}) and the related
expression of $\gamma$ in Eq.~(\ref{eq:dilemma})
well account for the balancing of 
gravitational and quantum forces.

In conclusion, we have obtained, 
in order of magnitude, the scales
of astrophysical and
cosmological structures through
a recursive geometric scheme. 
We have derived a sequence
of aggregates
starting from the minimum
scale (the nucleon) and then
moving upwards (bottom--up), by
iterating the alternating mechanisms of
tighter 3--d critical close packings 
and of looser 2--d close packings.
However, the sequence exhibits an
evident symmetry, in the
sense that it can be 
obtained travelling downwards (top--down).
Actually, it is possible, and moreover
likely, that the Universe started
from a 2--d close packing at low space energy
density on the
maximum scale $R$ and, due to some
fluctuative phenomena, it collapsed
onto critical 3--d close--packed 
structures on smaller
scales, later reaching a less confined 2--d 
close--packed configuration, and so on down to
the neutron stars.
The characterization of the 
temporal sequence, at variance with the
geometric one, remains an open question.

Looking forward to future developments,
the first aim is obviously
to improve the scheme introducing
corrections providing a more accurate
description of cosmological structures.
At the same time, we point out that there 
exist hints suggesting the possible
applicability to other, nongravitational
systems of mechanisms similar to the 
one presented here in the cosmological
context \cite{noiphysica}.

\end{document}